\documentclass[11pt,letter]{article}
\usepackage{geometry}                
\usepackage{amssymb}
\usepackage{epstopdf}

\usepackage[sf,bf]{caption}

\usepackage{gensymb}

\usepackage{graphics,graphicx}

\title{The cometary composition of a protoplanetary disk as revealed by complex cyanides\footnote{Manuscript accepted for publication in Nature. Reference: Nature, 520, 7546, 198, 2015}}

\author{Karin I. \"Oberg$^1$, Viviana V. Guzm\'an$^1$, Kenji Furuya$^2$, Chunhua Qi$^1$, Yuri Aikawa$^3$, \\Sean M. Andrews$^1$, Ryan Loomis$^1$, David J. Wilner$^1$}

\begin{document}
\maketitle

\begin{center}
$^1$Harvard-Smithsonian Center for Astrophysics, 60 Garden Street, Cambridge, MA 02138, USA.
 $^2$Leiden Observatory, Leiden University, P.O. Box 9513, 2300 CA Leiden, the Netherlands.
 $^3$Kobe University, 1-1 Rokkodaicho, Nada Ward, Kobe, Hyogo Prefecture 657-0013, Japan.
\end{center}

\begin{abstract}
Observations of comets and asteroids show that the Solar Nebula that spawned 
our planetary system was rich in water and organic molecules. Bombardment 
brought these organics to the young Earth's surface, seeding its early 
chemistry\cite{Hartogh11}. Unlike asteroids, comets preserve a nearly pristine record of the Solar Nebula composition.  The presence of cyanides in comets,
including 0.01\% of methyl cyanide (CH$_3$CN) with respect to water, is of special 
interest because of the importance of C-N bonds for abiotic amino acid
synthesis\cite{Goldman10}. Comet-like compositions of simple and complex volatiles are found in protostars, and can be readily explained by a
combination of gas-phase chemistry to form e.g. HCN and an active ice-phase 
chemistry on grain surfaces that advances complexity\cite{Herbst09}. 
Simple volatiles, including water and HCN, have been detected previously in 
Solar Nebula analogues -- protoplanetary disks around young stars -- indicating 
that they survive disk formation or are reformed 
{\it in situ}\cite{Dutrey97,Carr08,Hogerheijde11,Cleeves14}. 
It has been hitherto unclear whether the same holds for more complex organic 
molecules outside of the Solar Nebula, since recent observations show a dramatic change in the chemistry  at the boundary between nascent envelopes and young disks due to accretion shocks\cite{Sakai14}. Here we report the detection of CH$_3$CN (and 
HCN and HC$_3$N) in the protoplanetary disk around the young star MWC 480. We find 
abundance ratios of these N-bearing organics in the gas-phase similar to 
comets, which suggests an even higher relative abundance of complex cyanides 
in the disk ice.  This implies that complex organics accompany simpler 
volatiles in protoplanetary disks, and that the rich organic chemistry of the 
Solar Nebula was not unique. 
\end{abstract}

MWC~480 is a Herbig Ae star with an estimated stellar mass of 1.8 Solar masses 
(M$_\odot$) \cite{Simon00} in the Taurus star forming region at a distance of 
140~pc. The star is surrounded by a 0.18$\pm$0.1~M$_\odot$  protoplanetary disk, 
an order of magnitude more massive disk than the Minimum Mass Solar Nebula\cite{Guilloteau11,Takami14}. 
Compared to disks around Solar-type stars, the MWC 480 disk is 
2--3 times warmer at a given radius\cite{Chiang97,dAlessio04} and exposed to 
orders of magnitude higher ultraviolet radiation levels. Despite these environmental differences the volatile composition in the MWC 480 disk appears largely similar to that of disks around Solar-type stars, except for lower abundance of cold (T$<$20~K) chemistry tracers in the outer disk \cite{Dutrey07,Oberg10c}. The inner disk chemistry of MWC 480 has not been studied, but Herbig Ae and T Tauri disks are observed to present different volatile compositions close to their stars\cite{Pontoppidan10}.

Using the Atacama Large Millimeter/submillimeter Array (ALMA), we detected 
two emission lines of CH$_3$CN from the MWC 480 protoplanetary disk (the 
14$_0-13_0$ line at 5$\sigma$, and the 14$_1-13_1$ line at 3.5$\sigma$).  
We also detected emission lines from the N-bearing carbon chain HC$_3$N, and 
from the 13-C isotopologue of HCN. We targeted the 13-C isotopologue rather 
than the more abundant H$^{12}$CN because the latter is optically thick and 
therefore a poor tracer of the HCN abundance. Figure 1 shows the spatially 
resolved line detections together with a dust continuum emission map, 
demonstrating the spatial coincidence between CH$_3$CN, HC$_3$N, HCN and 
dust emission from the disk. The angular resolution is $0."4-0".6$, 
corresponding to 50--70~AU (1 AU is the distance from the Earth to the Sun). The emission is also spectrally resolved, which can be used to probe smaller spatial scales. Figure 1 shows the 
velocity gradient across the disk that arises from Keplerian rotation in all 
three lines.  Figure 2 shows the spectra of the three lines; each displays the 
double-peaked structure typical of a rotating disk.  Table 1 lists the 
integrated line fluxes. 

We use the spectrally and spatially resolved line emission to constrain the 
radial profiles of molecular column density based on parametric abundance models 
defined with respect to the adopted MWC 480 disk density and temperature 
structure (Methods). Figures 2 and 3a-c show the synthetic line spectra and 
maps from the best-fit models, demonstrating the good match between the models
and data. Figures 3d-f shows the best-fit radial column density profile, 
together with all profiles consistent with the data within 3$\sigma$, and abundances at 30 and 100~AU are reported in Table 1. 
The best-fit profiles have different slopes for the different molecules. 
The H$^{13}$CN column density decreases with radius, which is consistent with predictions from disk chemistry 
models\cite{Walsh10}. The increasing column density with radius of HC$_3$N, effectively a ring, is not predicted by models\cite{Walsh14}, indicative of that disk chemistry models
are incomplete for HC$_3$N. The CH$_3$CN emission is best reproduced with a flat profile, but other profiles cannot be excluded. 

The absolute abundances depend on the assumed disk density structure, and we therefore compare the complex cyanides in the MWC 480 disk to comet and 
to protostellar compositions using HCN as a reference species, similar to the practice in cometary studies\cite{Mumma11}. We calculate abundances of CH$_3$CN and HC$_3$N with 
respect to HCN at 30 AU, the smallest disk radius accessible by the ALMA
observations, and at 100~AU, the outer boundary of the cyanide emission maps.
Accounting for the higher luminosity of MWC 480 compared to the young Sun,
this radial range in the MWC 480 disk corresponds to the comet forming zone 
of 10--30~AU in the Solar Nebula\cite{Mumma11}. Assuming a standard 
HCN/H$^{13}$CN ratio of 70, the best-fit HC$_3$N and CH$_3$CN abundances with 
respect to HCN are 0.4 and 0.05 at 30~AU, and 5 and 0.2 at 100~AU, 
respectively. The CH$_3$CN/HCN abundance ratios are robust to model assumptions to within factors of a few, while the HC$_3$N/HCN 
abundance ratio may be overestimated by an order of magnitude using our simple abundance model (Methods). 
Conservatively, both the CH$_3$CN and HC$_3$N abundances with respect to HCN 
are thus $\sim$5\% at 30~AU and $\sim$20\% at 100~AU. A typical comet contains 
10\% of CH$_3$CN and HC$_3$N with respect to HCN\cite{Mumma11}. The MWC 480 
gas-phase cyanide composition at both 30 and 100~AU is thus cometary within the observational and model uncertainties.

The relationship between gas-phase abundance ratios and the abundance ratios in ices, the main volatile reservoirs in 
disks\cite{Semenov11,Walsh10, Walsh14}, depends on both desorption characteristics and chemistry (Methods). HCN, HC$_3$N and CH$_3$CN  are 
characterized by similar freeze-out and desorption kinetics, but different chemical pathways. In particular, the existence of efficient grain surface formation pathways to CH$_3$CN enhances CH$_3$CN 
with respect to the other cyanides in the ice mantles. The scale 
of this enhancement factor varies among models, but it is at least one order 
of magnitude (Methods). This results in an expected minimum CH$_3$CN/HCN ice ratio of 
0.5 at 30~AU in the MWC 480 disk, considerably higher than the typical Solar System comet.

The CH$_3$CN/HCN and HC$_3$N/HCN ratios in the MWC 480 disk are also high 
when compared to protostars. 
The HCN/HC$_3$N/CH$_3$CN ratio is 1/0.01/0.08 toward the Solar-type 
protostellar binary IRAS 16298-2422\cite{vanDishoeck95}, and similar abundance 
ratios are found toward more massive systems. The MWC 480 cyanide composition is thus difficult to explain by inheritance alone, as has been suggested for e.g. 
H$_2$O\cite{Cleeves14}. Rather the observed high disk abundances likely reflect an efficient 
disk chemistry that readily converts a large portion of the carbon originally 
in CO and other small molecules into more complex 
organics\cite{Ciesla12,Favre13} during the first million years of the disk 
life time. 

This early, efficient complex chemistry in protoplanetary disks impacts the surface conditions of rocky (exo-)planets. In the Grand Tack model 
of early Solar System dynamics\cite{Tsiganis05}, the coupled migration of Jupiter and Saturn in the gas-rich disk from which they formed caused scattering of volatile-rich (and thus complex 
organic-rich) planetesimals inwards, mixing with the asteroid belt, and 
outwards, producing the present-day Kuiper belt and comets\cite{Walsh11}. While comets are expected (and observed) to conserve most of their original compositions due to low temperatures and thus very long desorption and chemical time scales\cite{Mumma11}, asteroid belt bodies are more processed. A combination of evaporation and chemistry probably resulted in both a net loss of volatiles and in a relative enhancement of complex organics with respect to H$_2$O over the subsequent 10s of Myr\cite{Pizzarello12}. A second instability 100 million years later resulted in a heavy bombardment 
of the Earth by mainly asteroid belt bodies, including the icy bodies originating 
in the outer Solar System\cite{Obrien14}.  The frequency of these instabilities 
in exo-planetary systems is unknown, but the pile-up of Hot Jupiters and 
Super-Earths reveal that major planetary migrations are common. When these
migrations cause bombardment of icy bodies, they are likely organic-rich.  

The high ratio of complex 
to simple cyanides in the MWC 480 disk implies that the rich organic 
composition of comets is not unique to our Solar System, and could be 
common-place. Laboratory experiments have shown that the same ice chemistry that produce CH$_3$CN, i.e. photo processing of interstellar ice analogs also produce simple sugars and amino acids\cite{Oberg09d,MunozCaro02}. This suggests that the early surface conditions of Earth, 
set by comet and asteroid bombardment, may be common for young rocky planets, 
and that conditions favourable to an even richer chemistry may be ubiquitous.


\subsection*{Addendum}
\begin{description}
 \item [Acknowledgements] The authors would like to acknowledge helpful comments from Ewine van Dishoeck. This paper makes use of ALMA data. ALMA is a partnership of ESO (representing
its member states), NSF (USA) and NINS (Japan), together with NRC
(Canada) and NSC and ASIAA (Taiwan), in cooperation with the Republic of
Chile. The Joint ALMA Observatory is operated by ESO, AUI/NRAO and NAOJ. The National Radio Astronomy Observatory is a facility of the National Science Foundation operated under cooperative agreement by Associated Universities, Inc. K.I.\"O. would like to especially acknowledge Adam Leroy and the NAASC for assistance with calibration and imaging. K.I.\"O. also acknowledges funding from the Simons Collaboration on the Origins of Life (SCOL), the Alfred P. Sloan Foundation, and the David and Lucile Packard Foundation. D.J.W. acknowledges funding from NASA Origins of Solar Systems grant No. NNX11AK63.
 \item[Competing Interests] The authors declare that they have no
competing financial interests.
 \item[Author contributions] K.I.\"O led the overall project, reduced the data, assisted by V.V.G. and R.L., and wrote the manuscript with revisions from S.M.A. and D.J.W. V.V.G., assisted by C.Q., performed the parametric modelling and abundance extraction. K.F. performed the astrochemical modelling, and interpreted the results with Y.A. All authors contributed to discussions of the results and commented on the manuscript.
 \item[Author information] Correspondence and requests for materials
should be addressed to K.I.\"O.~(email: koberg@cfa.harvard.edu). The ALMA program number for the presented data is 2013.1.00226.
\end{description}

\begin{table}
\centering
\caption{Molecular data.}
\medskip
\begin{tabular}{lccccccc}
\hline
Molecule & QN   & Rest freq.    &E$_{\rm u}$   & Integrated  &$x_{\rm 30 AU}^{\rm a}$ &$x_{\rm 100 AU}^{\rm a}$\\
&&&&flux [rms]\\
 &    & GHz    & K    &mJy km/s &($10^{-13}$ $n_{\rm H}^{-1}$)&($10^{-13}$ $n_{\rm H}^{-1}$)\\
 &&&&beam$^{-1}$\\
\hline
H$^{13}$CN	&J=3--2				&259.0118	&24.9		&47[8]	&1.6 [1.3--2.0]	&1.6 [1.3--2.0]\\
HC$_{3}$N	&J=27--26				&245.6063	&165			&66[8]	&43 [36--54]&480 [400--600]\\
CH$_3$CN	&$14_{\rm0}-13_{\rm0}$	&257.5274	&92.7		&30[6]	&4.8 [2.3--7.5]	&16 [6--55]\\
CH$_3$CN	&$14_{\rm1}-13_{\rm1}$	&257.5224	&100			&23[6]	&\\
\hline
\end{tabular}
$^{\rm a}$best-fit abundances assuming a vertically constant abundance. the 3$\sigma$ abundance range is in [].
\end{table}

\begin{figure}
\includegraphics[width=1.0\textwidth]{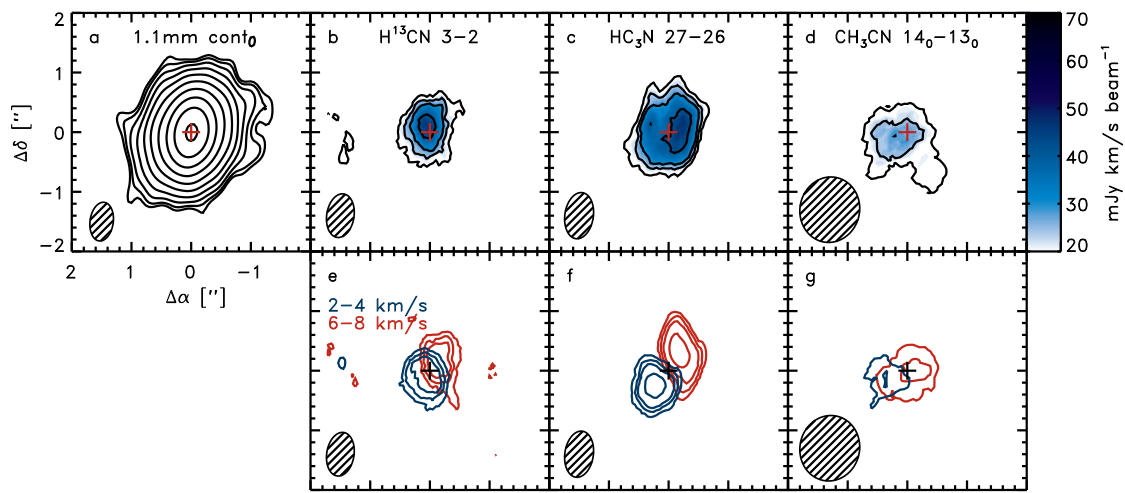}
\caption{\textbf{ALMA detections of simple and complex cyanides in the MWC 480 protoplanetary disk.} 
a, 1.1 mm emission (black contours are 3$\sigma+\sigma\times2^{[1,2,...]}$). 
b--d, integrated emission of H$^{13}$CN (a), HC$_3$N (b) and CH$_3$CN (c) lines (colour: see colour scale on the right). Black contours are [3,4,5,7,10]$\sigma$. e--g, as b--d, but for 2~km/s velocity bins around the source mean velocity, displaying the disk rotation. Positions are relative to the continuum phase centre (marked with a cross) at right ascension (RA) 04:58:45.94 and declination (DEC) +29:50:38.4. The synthesized beam is shown in the bottom left corner of each panel.}
\end{figure}

\begin{figure}
\includegraphics[width=0.5\textwidth]{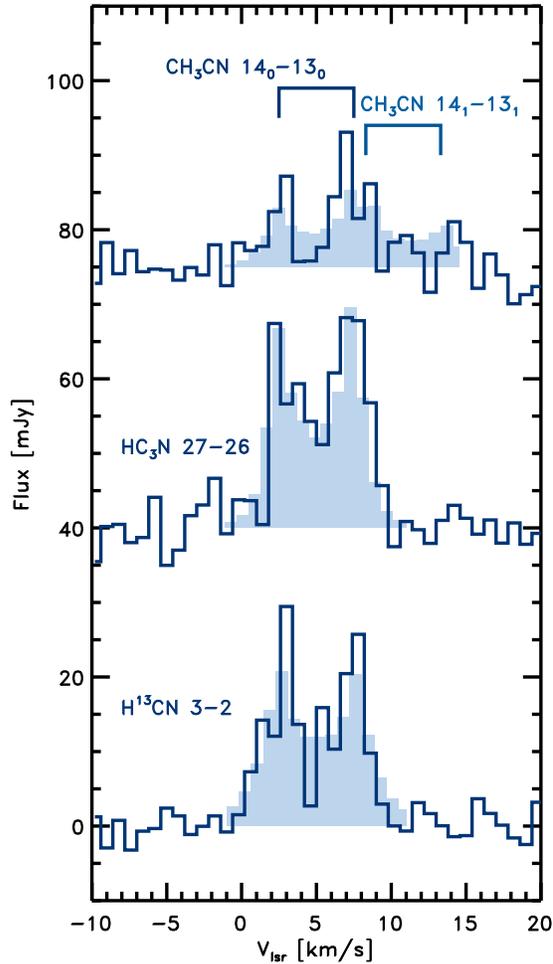}
\caption{\textbf{Spectra of detected cyanides in the MWC 480 protoplanetary disk.} The observed spectra (contours) of H$^{13}$CN, HC$_3$N, and CH$_3$CN are extracted from ALMA spectral-image data cubes. The synthetic spectra (histograms) are based on the best fit disk abundance models in Fig. 3. The CH$_3$CN spectrum contains two partially overlapping lines identified with the 14$_0-13_0$ and 14$_1-13_1$ transitions. The spectra were extracted from the spectral image cubes using a Keplerian mask to maximize the signal-to-noise. }
\end{figure}

\begin{figure}
\includegraphics[width=0.8\textwidth]{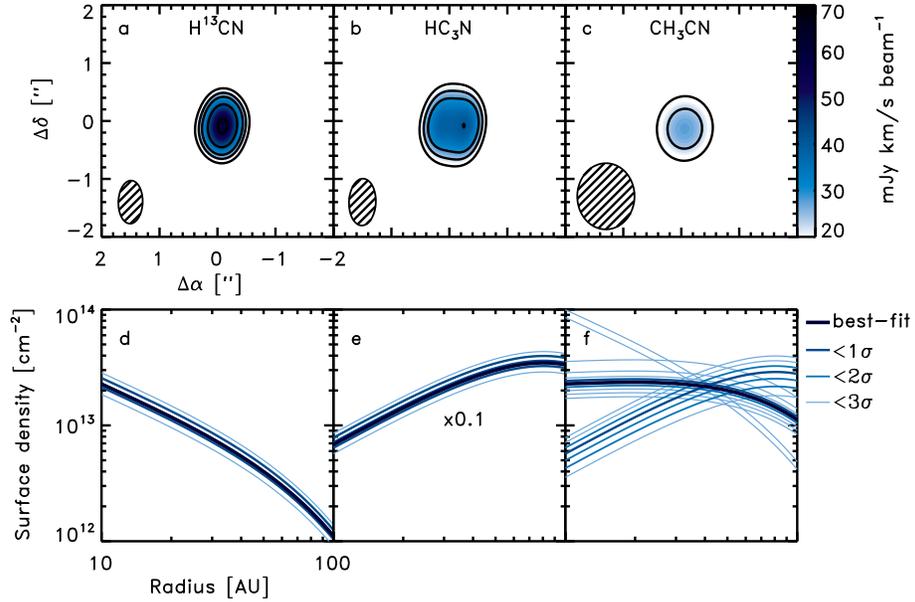}
\caption{\textbf{Models of cyanide emission and radial distributions in the MWC 480 disk.} a-c, synthetic ALMA observations without added noise based on the best-fit radial column density profiles of H$^{13}$CN, HC$_3$N, and CH$_3$CN (colour: see colour scale on the right). Black contours are as in Fig. 1. d-e, best-fit radial column density profiles (thick dark blue line), shown together with all column density profiles that are consistent with data within 1, 2 and 3$\sigma$ confidence intervals. Increasing deviations from the best-fit are shown with successively lighter shades and thinner lines.}
\end{figure}

\newpage
\section*{Methods}

\subsection*{ALMA observations and data reduction.}

MWC 480 was observed with the Atacama Large Millimeter/submillimeter Array as 
a part of the Cycle 2 proposal 2013.1.00226. Observations were carried 
out on 2014 June 15, with 33 antennas and baselines of 18--650~m 
(15--555~k$\lambda$). The total on source integration time was 22 minutes. 
The nearby quasar J0510+1800 was used for phase and gain, bandpass 
 and absolute flux calibration (J0510+1800 was itself 
calibrated to an absolute flux of 1.50[0.08] Jy on 2014 June 29). The correlator was configured to observe 14 spectral windows (SPWs) with 
resolution $\delta\nu$= 61.04 kHz.  The H$^{13}$CN 3--2 line was placed in 
SPW 1, HC$_3$N 27-26 in SPW 9 and CH$_3$CN 14$_{\rm 0}$-13$_{0}$ and 
14$_{\rm 1}$-13$_{1}$ in SPW 5, each 60--120~MHz wide and centered on 
258.983, 245.622 and 257.498 GHz, respectively. 

The visibility data were calibrated by ALMA/ NAASC staff following standard 
procedures. Each SPW was further self-calibrated
in phase and amplitude in CASA 4.2.2. The self-calibrated phase and gain 
SPW-specific solutions were applied to create individual SPW line data cubes
using the Briggs weighting scheme, with Brigg's robustness parameter set to 1--2.
For the weak CH$_3$CN lines, a taper of 1$''$ was applied to the visibilties 
to maximize the signal-to-noise.
The data cubes were subsequently continuum subtracted using channels
identified as free of spectral line emission. The resulting rms noise in 
1 km/s bins was $\sim$2.3~mJy per beam. A CLEAN deconvolution was performed with a 
mask produced manually based on the H$^{13}$CN emission and applied to the 
HC$_3$N and CH$_3$CN spectral line cubes (after verifying that the HC$_3$N 
and CH$_3$CN cubes contained no additional emission).
For the 257~GHz continuum map, obtained by combining all SPWs in the USB),
the synthesized beam is $0.''65 \times 0.''39$ (PA=-7\degree) and rms noise 
0.3~mJy beam$^{-1}$.  The 257 GHz continuum flux density was determined to be 
331$\pm$33~mJy (assuming a 10\% uncertainty in the absolute calibration), 
which agrees well with previous Submillimer Array results$^{\rm 30}$.  

Integrated emission maps were produced using the $immoments$ task in CASA, 
adding emission $>$1$\sigma$ from each channel. Lower clipping levels 
do not change the result at the 10\% level. Spectra were extracted using both 
the empirical CLEAN masks and theoretical masks based on Keplerian rotation
to isolate disk emission in each channel. The CLEAN masks resulted in the 
best signal-to-noise ratio, and these results are used in the study.

\subsection*{Parametric disk density and temperature model.}

We adopt a previously developed disk density structure of 
MWC 480\cite{Guilloteau11}. The total surface density is given by the 
similarity solution, which self-consistently describes the density structure 
of an accretion disk$^{31,32}$:

\begin{equation}
  \Sigma(r) = \Sigma_c \left (\frac{r}{R_c}\right)^{-\gamma} \exp
  \left ( -\left ( \frac{r}{R_c}\right )^{2-\gamma}\right ),
\end{equation}

\noindent where $\Sigma_c=M_{gas}(2-\gamma)/(2 \pi R^2_c)$ is a normalization coefficient,
$R_c$ is the characteristic radius, and $\gamma$ is a gradient parameter 
describing the density fall-off with radius. The two-dimensional density 
structure in cylindrical coordinates is 

\begin{equation}
  \rho(r,z) = \frac{\Sigma(r)}{H(r)\pi} \exp \left (\frac{-z^2}{H(r)^2} \right),
\end{equation}

\noindent where $H(r)=H_{100}(r/100AU)^h$ is the scale height, and $h$ a parameter that
describes the amount of flaring of the disk. The density structure parameters 
were constrained using high angular resolution observations of the MWC 480 
dust continuum emission at 1.3 mm and 3 mm wavelengths \cite{Guilloteau11} 
and are listed in Extended Data Table 1. The resulting mass surface densities in g cm$^{-2}$ are $550$,  $91$ and $4.7$ at 1, 10 and 100 AU, respectively. 

Based on existing model fits to dust and $^{13}$CO observations$^{33,34}$, the disk 
midplane temperature is parametrized as 
$T(r) = T_0 \left( \frac{r}{R_0}\right)^{-q}$, where $T_0$ is set to 23~K at 
$R_0=100$~AU, and the power-law index $q$ is 0.4. To account for the presence 
of a vertical temperature gradient due to heating of the disk surface by the 
central star$^{34,35}$, we 
parameterize the two-dimensional temperature structure in cylindrical 
coordinates as 

\begin{equation}
  T(r,z) = T_0 \left( \frac{r}{R_0}\right)^{-q} \exp \left(\log \beta \frac{z}{H(r)} \right),
\end{equation}

\noindent where $T_0$ is the midplane temperature at $R_0$, $q$ is a power-law index 
describing the decrease in midplane temperature with radius, and $\beta$
is a factor describing the increase in temperature at increasing disk height, 
which is set to 1.5$^{\rm 34}$. The density and temperature disk 
structures are shown in Extended Data Fig. 1. 

\subsection*{Molecular abundance retrieval.}

To retrieve molecular abundance and column density profiles, we define 
parametric abundance models with respect to the adopted MWC 480 disk density 
and temperature structure. We use a simple power-law prescription 
$x=x_{\rm 100AU}\times(r/100~AU)^\alpha$ for the molecular abundances where 
$x$ is the abundance of the molecule with respect to the total hydrogen 
density, $x_{\rm 100AU}$ is the abundance at 100 AU, $r$ is the disk radius in 
AU, and $\alpha$ is a power law index$^{36}$. To obtain a column density 
profile, which is more intuitively related to observations, the abundance 
prescription is multiplied by disk surface density profile, Eq. 1. We set an 
outer cut-off radius of 100~AU, corresponding to the most extended emission 
observed for any of the cyanides.  We also explored $R_{\rm out}$ of 60 and 
80~AU, but this did not improve the fit for any of the molecules.  For each 
molecule, we calculate a grid of abundance models that covers $x_{\rm 100AU}$ 
from $10^{-14}-10^{-10}$ with respect to $n_{\rm H}$, and $\alpha$ between 0 
and 1 for H$^{13}$CN, 1 and 2 for HC$_3$N, 0 and 2 for CH$_3$CN.  These 
initial ranges of $\alpha$ were selected based on visual inspection of the 
observed emission, noting its level of central concentration, i.e. 
$\alpha=0$ corresponds to a steeply decreasing column density profile, 
$\alpha=1$ to an almost flat profile, and $\alpha=2$ to an increasing 
profile with radius.

The best-fit models are obtained by minimizing $\chi^2$, the weighted
difference between the real and imaginary part of the complex visibility 
measured in the $(u,v)$-plane sampled by the ALMA observations$^{36,37,38}$. We use the three-dimensional Monte Carlo code LIME$^{39}$  to calculate the radiative transfer and molecular excitation. 
In addition to the molecular abundance profile and the disk density and 
temperature structure parameters, the radiative transfer modelling requires 
information on disk inclination, turbulence and velocity field. We adopted a 
disk inclination of 37$^{\circ}$, disk turbulence of 0.05 km/s and a Keplerian 
velocity field with a stellar mass of 1.8 M$_\odot$\cite{Guilloteau11}. 
For HCN and HC$_3$N, the level populations were computed in non-LTE (Local Thermodynamic Equilibrium) with 
collision rates listed in the BASECOL 
database$^{40,41}$.  For CH$_3$CN, we assumed LTE, which 
is a reasonable approximation, since the expected disk emissive layers have 
higher densities than the critical density. We checked this assumption by 
running several HCN and HC$_3$N models assuming LTE and found a good agreement 
(within 20\%) between LTE and non-LTE models. In Extended Data Fig. 2 we show 
the best-fit models for different $\alpha$. In the case of HC$_3$N, only the 
models with $\alpha=2$ shows the observed ring-like structure and are 
therefore the only ones considered further.  

Within the adopted model framework, the molecular abundances and abundance 
ratios are constrained within a factor of a few. These models implicitly 
assume a constant abundance profile with disk height, which may be a poor 
approximation in the outer disk where freeze-out in the midplane results in a 
vertically layered structure\cite{Semenov11}$^{,42,43}$. The existing 
data are, however, insufficient to put any constraints on the vertical 
distribution. Rather than pursuing a more complex model, we opt to test the 
sensitivity of the derived abundances and abundance ratios on the details of 
the vertical structure assumptions. We ran a second grid of models to 
simulate the effect of freeze-out in the midplane by reducing the molecular 
abundances by three orders of magnitude at $z/R<0.2$. Using the same fitting 
procedure as before, we find best-fit  $x_{\rm 100 au}$ of 
$1\times10^{-11}$ ($\alpha$=0), $3.5\times10^{-10}$ ($\alpha$=2),  
and $5\times10^{-11}$ ($\alpha$=1), for H$^{13}$CN, HC$_3$N and CH$_3$CN, 
respectively. These abundances are an order of magnitude higher than those 
obtained assuming a vertically constant abundance. The CH$_3$CN/H$^{13}$CN 
abundance ratio profile is unchanged within the model uncertainties, however. Despite a higher excitation level of CH$_3$CN the two molecules emit from a similar disk layer when assigned the same vertical abundance profile due to a combination of a high critical density of the CH$_3$CN transition and a rapid fall-off in density with disk height, i.e. practically no CH$_3$CN emission originates at $z/R>0.5$. In contrast, the derived HC$_3$N/HCN ratio does depend on the 
vertical structure. The HC$_3$N transition has an even higher energy level 
and a lower critical density than CH$_3$CN, which implies that some of the 
HC$_3$N emission can originate in more elevated disk layers than H$^{13}$CN 
and CH$_3$CN. The reported CH$_3$CN/H$^{13}$CN results are thus robust to the abundance model assumptions, while the reported
HC$_3$N/H$^{13}$CN results may be overestimated by up to an order magnitude.

\subsection*{Disk chemistry modelling: constraining the ice-to-gas ratios.}

To constrain the origin of CH$_3$CN and therefore how its ice-to-gas ratio compares with HCN, we first explore how much of the observed CH$_3$CN gas could come from gas-phase chemistry. We ran a grid of pseudo-time dependent models with a complete gas-phase chemistry$^{44}$, but without any grain surface 
reactions (except for adsorption and desorption, and H$_2$ formation on grains), for 1 Myr. The grid covers several orders of magnitude of densities, temperatures, UV fields and ionization fractions -- the four most important regulators of gas-phase chemistry in disks. For gas-phase CH$_3$CN production, the ionization rate is expected to be of special importance because the main gas-phase formation pathway includes CH$_3^+$ (through HCN+CH$_3^+$, which has a expert-validated reaction rate in the KIDA database$^{45}$). Extended Data Figure 3 shows that none of these models can produce CH$_3$CN/HCN $>$0.01 and that in the vast majority of the parameter space CH$_3$CN/HCN$<$0.001. A ratio of 0.01 is approached for an ionization rate of 10$^{-14}$ s$^{-1}$, but such high ionization rates are only attained in the disk atmosphere where stellar X-rays regulate the ionization balance. This disk layer contain a very small fraction of the total disk mass and therefore contributes a negligible amount of the total molecular column. Deeper into the disk, the ionization rate is reduced below 10$^{-17}$ s$^{-1}$ because of attenuation of both X-rays and cosmic rays$^{46}$. A ratio of 0.01 is also approached at lower ionization rates at very low temperatures and high densities. Such environments are characteristic for the outer disk midplane, which contain a lot of mass, but freeze-out at these temperatures results in CH$_3$CN abundances below 10$^{-14}$, which would not contribute to the observed CH$_3$CN emission. The observed high abundance of CH$_3$CN gas with respect to HCN therefore implies that grain surface 
chemistry contributes significantly to the observed CH$_3$CN abundance.

We ran a number of complete disk chemistry models for 1Myr, the estimated age of the MWC 480 star+disk system, which include gas and grain 
surface chemistry, as well as different levels of turbulence, to calculate the 
range of plausible ice-to-gas conversion factors for HCN (and other gas-phase 
chemistry products) and CH$_3$CN$^{47}$. The main grain surface CH$_3$CN formation pathways in these models are hydrogenation of C$_2$N and CH$_3$+CN and ice photochemistry. The main desorption pathway is UV photodesorption because of the high binding energies of HCN. HC$_3$N and CH$_3$CN: 4170, 4580, and 4680~K, respectively$^{48,49,50}$, respectively, corresponding to sublimation temperatures of 
 $\sim$90~K at a gas density of 
10$^6$ cm$^{-3}$ and 110~K at the gas density of 10$^{10}$ cm$^{-3}$. The three cyanides should also present very similar UV photodesorption efficiencies because of similar radiation cross-sections and binding energies$^{51}$. Absent grain surface chemistry, these molecules would thus have had comparable ice-to-gas ratios.

Because CH$_3$CN can form on the grains in the cold midplane a key feature is the coupling between the ice formation layer and the disk layers where molecules are efficiently desorbed into the gas-phase. Without such a coupling, ice chemistry cannot affect the gas-phase volatile composition.
In the disk models, we define the vertical diffusion coefficient as $D_z = \alpha_z c_s^2/\Omega$, where $\alpha_z$ is a free parameter, $c_s$ is the local sound velocity, and $\Omega$ is the Keplerian orbital frequency. Different levels of turbulence are simulated by different values of $\alpha_z$ between 0 and 0.01. 
In models with turbulence turned off ($\alpha_z=0$), there is no mixing 
between the cold, icy midplane and the intermediate layers where gas-phase 
molecules are abundant (and the observed cyanides likely reside). Extended Data Figure 4a shows that without turbulence the predicted CH$_3$CN abundances are low everywhere ($<10^{-11}$ w.r.t $n_{\rm H}$, where $n_{\rm H}$ is the number density of H nuclei) and the vertically averaged CH$_3$CN/HCN abundance ratio is $<$0.001; see Extended Data Figure 5a,d).

In models that include vertical mixing  ($\alpha_z=0.001-0.01$), the CH$_3$CN abundance as well as the CH$_3$CN/HCN abundance ratio are enhanced by an order of magnitude because of mixing of midplane icy grains into UV exposed disk regions where the ices photodesorb (Extended Data Figs. 4 and 5). In particular the CH$_3$CN/HCN ratio approaches 10$^{-2}$ outside of 30~AU, within an order of magnitude of the observed ratios. Model-data comparison thus suggest that the intermediate disk 
layers are strongly coupled to the icy midplane through mixing. This implies 
that gas-phase observations of CH$_3$CN and other cyanides at intermediate 
disk heights can be used to trace the total volatile reservoir, even if the 
conversion from gas to ice can be complex.

Using the three models, with no, moderate and strong mixing we calculate the 
HCN and CH$_3$CN ice-to-gas density and column density ratios as a function of disk radius 
(Extended Data Fig. 6,7). For HCN, the ice-to-gas ratio varies between 
10$^5$ and 10$^3$ between 30 and 100~AU. For CH$_3$CN the 
ice-to-gas ratio varies between 10$^7$ and 10$^4$. In other disk chemistry 
models\cite{Walsh14}$^{,52}$ the contrast between the two ice-to-gas ratios 
is even higher. Considering the range of values and their sensitive dependence 
on assumptions about disk turbulence, we conclude that 
the ice-to-gas ratio for CH$_3$CN is at least 10 times higher than HCN and 
HC$_3$N. At longer time scales the balance between formation in the midplane and destruction at more exposed disk layers determines whether the CH$_3$CN abundance increases or decreases -- in general the more turbulence the faster the destruction. 

It is important to note, that we employ generic disk structures$^{53}$ that have not been fit to the MWC 480 data. The temperature and density structures are comparable (cf. Extended Data Figs. 1 and 4). The UV field is lower than would be expected toward a Herbig Ae star and the gas-phase CH$_3$CN production may therefore be somewhat overestimated in our model, but as shown above, gas-phase chemistry is a minor contributor to the overall CH$_3$CN gas budget. The grain surface production takes place in UV shielded regions and should thus not be affected by an increased UV flux. An increased UV field intensity may  change the desorption/adsorption balance, however, since UV photodesorption is an important desorption mechanism. This increase in photodesorption will be accompanied by an increased level photodissociation and the main effect will therefore be to push the desorbing layer closer to the midplane rather than increasing its abundance. An A star type radiation field may aid the detection of gas-phase CH$_3$CN, since less turbulence may be required to bring the icy grains up to UV exposure and into the gas-phase. A decrease in the disk height of the cyanide gas layer, compared to our generic disk model, may also reduce the absolute ice-to-gas ratios, but the relative ratios of species with similar adsorption and desorption characteristics should not be significantly affected.

\subsection*{Desorption Mechanism}

In our disk chemistry model, UV photodesorption regulates the desorption of oh HCN, HC$_3$N and CH$_3$CN ice into the gas-phase. This result depends on a combination of disk and star properties and on the theoretical yields of different desorption mechanisms, many of which are poorly constrained. While relative ice-to-gas ratios of related species are insensitive to the details of the desorption process, the absolute ice-to-gas ratios found in our models must be considered highly uncertain. Based on theory, different desorption mechanisms should present different radial and vertical abundance distributions, which could be tested observationally. Desorption due to release of chemical formation energy should be most important in the cold disk midplane, while UV photodesorption is only efficient in warmer, upper disk layers, and thermal desorption in the disk layers that are warm enough. In particular, colder molecular emission layers should be characterized by lower excitation temperatures, which can be measured when multiple lines of the same species are observed. Observational constraints on ice desorption in disks will thus be attainable with ALMA.

Different desorption mechanisms should also exhibit different chemical signatures, although the details are often model-dependent.  In the case of thermal desorption, the gas phase composition in a layer should simply depend on volatility, albeit with the added complication of ice entrapment$^{54}$. Chemistry can also help distinguish between ice and gas origins and thus provide an independent measure of the relative importance of the two pathways. he high ratio of CH$_3$CN/CH$_3$NC in the Horsehead PDR (Photon Dominated Region), for example, was used as evidence for a grain surface origin of the observed CH$_3$CN$^{55}$. Chemical tracers could also provide useful constraints on the incident radiation fields and temperature structure and thus aid in ruling out some of the potentially important desorption mechanisms. For example CN/HCN is a proposed tracer of UV flux and HNC/HCN of temperature$^{56,57}$. \\

\begin{enumerate}
  \setcounter{enumi}{29}
  \item {\"O}berg, K.~I. \emph{et~al.} {Disk Imaging Survey of Chemistry with SMA. II.
  Southern Sky Protoplanetary Disk Data and Full Sample Statistics}. \emph{{Astrophys. J.}} \textbf{734}, {98} ({2011}).
  
  \item {{Lynden-Bell}, D.} \& {{Pringle}, J.~E.} {{The evolution of viscous discs and the origin of the nebular variables.}}
\newblock \emph{{Mon. Not. R. Astron. Soc.}} \textbf{{168}}, {603--637} ({1974}).

\item {{Hartmann}, L.}, {{Calvet}, N.}, {{Gullbring}, E.} \& {{D'Alessio}, P.}
\newblock {{Accretion and the Evolution of T Tauri Disks}}.
\newblock \emph{{Astrophys. J.}} \textbf{{495}},
  {385--400} ({1998}).
  
  \item {{Pi{\'e}tu}, V.}, {{Dutrey}, A.},
  {{Guilloteau}, S.}, {{Chapillon}, E.} \&
  {{Pety}, J.}
\newblock {{Resolving the inner dust disks surrounding LkCa 15
  and MWC 480 at mm wavelengths}}.
\newblock \emph{{Astron. Astrophys.}} \textbf{{460}},
  {L43--L47} ({2006}).
  
  \item {{Dartois}, E.}, {{Dutrey}, A.} \&
  {{Guilloteau}, S.}
\newblock {{Structure of the DM Tau Outer Disk: Probing the
  vertical kinetic temperature gradient}}.
\newblock \emph{{Astron. Astrophys.}} \textbf{{399}},
  {773--787} ({2003}).
  
  \item {{Rosenfeld}, K.~A.}, {{Andrews}, S.~M.},
  {{Wilner}, D.~J.}, {{Kastner}, J.~H.} \&
  {{McClure}, M.~K.}
\newblock {{The Structure of the Evolved Circumbinary Disk
  around V4046 Sgr}}.
\newblock \emph{{Astrophys. J.}} \textbf{{775}},
  {136} ({2013}).
  
  \item {{Qi}, C.} \emph{et~al.}
\newblock {{Imaging of the CO Snow Line in a Solar Nebula
  Analog}}.
\newblock \emph{{Science}} \textbf{{341}},
  {630--632} ({2013}).
  
  \item {{Qi}, C.}, {{Wilner}, D.~J.},
  {{Aikawa}, Y.}, {{Blake}, G.~A.} \&
  {{Hogerheijde}, M.~R.}
\newblock {{Resolving the Chemistry in the Disk of TW Hydrae. I.
  Deuterated Species}}.
\newblock \emph{{Astrophys. J.}} \textbf{{681}},
  {1396--1407} ({2008}).
  
  \item {{Qi}, C.} \emph{et~al.}
\newblock {{Resolving the CO Snow Line in the Disk around HD
  163296}}.
\newblock \emph{{Astrophys. J.}} \textbf{{740}},
  {84} ({2011}).
  
  \item {{Brinch}, C.} \& {{Hogerheijde}, M.~R.}
\newblock {{LIME - a flexible, non-LTE line excitation and
  radiation transfer method for millimeter and far-infrared wavelengths}}.
\newblock \emph{{Astron. Astrophys.}} \textbf{{523}},
  {A25} ({2010}).
  
  \item {{Wernli}, M.}, {{Wiesenfeld}, L.},
  {{Faure}, A.} \& {{Valiron}, P.}
\newblock {{Rotational excitation of HC\_3N by H$_{2}$ and He at
  low temperatures}}.
\newblock \emph{{Astron. Astrophys.}} \textbf{{464}},
  {1147--1154} ({2007}).

\item {{Dubernet}, M.}, {{Nenadovic}, L.} \&
  {{Doronin}, N.}
\newblock {{SPECTCOL: A New Software to Combine Spectroscopic
  Data and Collisional Data within VAMDC}}.
\newblock In {{Ballester}, P.}, {{Egret}, D.}
  \& {{Lorente}, N.~P.~F.} (eds.)
  \emph{{Astronomical Data Analysis Software and Systems XXI}}, vol. {461} of \emph{{Astronomical
  Society of the Pacific Conference Series}}, {335--338}
  ({2012}).
  
\item {{Aikawa}, Y.} \& {{Herbst}, E.}
\newblock {{Molecular evolution in protoplanetary disks.
  Two-dimensional distributions and column densities of gaseous molecules}}.
\newblock \emph{{Astron. Astrophys.}} \textbf{{351}},
  {233--246} ({1999}).
  
\item {{Walsh}, C.}, {{Millar}, T.~J.} \&
  {{Nomura}, H.}
\newblock {{Molecular Line Emission from a Protoplanetary Disk
  Irradiated Externally by a Nearby Massive Star}}.
\newblock \emph{{Astrophys. J.L}} \textbf{{766}},
  {L23} ({2013}).
  
  \item {{Aikawa}, Y.}, {{Wakelam}, V.},
  {{Hersant}, F.}, {{Garrod}, R.~T.} \&
  {{Herbst}, E.}
\newblock {{From Prestellar to Protostellar Cores. II. Time
  Dependence and Deuterium Fractionation}}.
\newblock \emph{{Astrophys. J.}} \textbf{{760}},
  {40} ({2012}).
  
\item {{Wakelam}, V.} \emph{et~al.}
\newblock {{A KInetic Database for Astrochemistry (KIDA)}}.
\newblock \emph{{Astrophys. J.S}} \textbf{{199}},
  {21} ({2012}).
  
  \item {{Cleeves}, L.~I.}, {{Bergin}, E.~A.},
  {{Qi}, C.}, {{Adams}, F.~C.} \&
  {{Oberg}, K.~I.}
\newblock {{Constraining the X-ray and Cosmic Ray Ionization
  Chemistry of the TW Hya Protoplanetary Disk: Evidence for a Sub-interstellar
  Cosmic Ray Rate}}.
\newblock \emph{{ArXiv e-prints}}  ({2014}).

\item {{Furuya}, K.} \& {{Aikawa}, Y.}
\newblock {{Reprocessing of Ices in Turbulent Protoplanetary
  Disks: Carbon and Nitrogen Chemistry}}.
\newblock \emph{{Astrophys. J.}} \textbf{{790}},
  {97} ({2014}).
  
\item {{Yamamoto}, T.}, {{Nakagawa}, N.} \&
  {{Fukui}, Y.}
\newblock {{The chemical composition and thermal history of the
  ice of a cometary nucleus}}.
\newblock \emph{{Astron. Astrophys.}} \textbf{{122}},
  {171--176} ({1983}).
  
\item {{Garrod}, R.~T.} \& {{Herbst}, E.}
\newblock {{Formation of methyl formate and other organic
  species in the warm-up phase of hot molecular cores}}.
\newblock \emph{{Astron. Astrophys.}} \textbf{{457}},
  {927--936} ({2006}).
  
\item {{Collings}, M.~P.} \emph{et~al.}
\newblock {{A laboratory survey of the thermal desorption of
  astrophysically relevant molecules}}.
\newblock \emph{{Mon. Not. R. Astron. Soc.}} \textbf{{354}},
  {1133--1140} ({2004}).
  
\item {van Dishoeck, E.~F.}, {Jonkheid, B.} \&
  {van Hemert, M.~C.}
\newblock {Photoprocesses in protoplanetary disks}.
\newblock \emph{{Faraday Discussions}}
  \textbf{{133}}, {231--343}
  ({2006}).
  
\item {{Walsh}, C.}, {{Nomura}, H.},
  {{Millar}, T.~J.} \& {{Aikawa}, Y.}
\newblock {{Chemical Processes in Protoplanetary Disks. II. On
  the Importance of Photochemistry and X-Ray Ionization}}.
\newblock \emph{{Astrophys. J.}} \textbf{{747}},
  {114} ({2012}).
  
\item {{Nomura}, H.}, {{Aikawa}, Y.},
  {{Tsujimoto}, M.}, {{Nakagawa}, Y.} \&
  {{Millar}, T.~J.}
\newblock {{Molecular Hydrogen Emission from Protoplanetary
  Disks. II. Effects of X-Ray Irradiation and Dust Evolution}}.
\newblock \emph{{Astrophys. J.}} \textbf{{661}},
  {334--353} ({2007}).
  
\item {{Fayolle}, E.~C.}, {{{\"O}berg}, K.~I.},
  {{Cuppen}, H.~M.}, {{Visser}, R.} \&
  {{Linnartz}, H.}
\newblock {{Laboratory H$_{2}$O:CO$_{2}$ ice desorption data:
  entrapment dependencies and its parameterization with an extended three-phase
  model}}.
\newblock \emph{{Astron. Astrophys.}} \textbf{{529}},
  {A74} ({2011}).
  
\item {{Gratier}, P.} \emph{et~al.}
\newblock {{The IRAM-30 m line survey of the Horsehead PDR. III.
  High abundance of complex (iso-)nitrile molecules in UV-illuminated gas}}.
\newblock \emph{{Astron. Astrophys.}} \textbf{{557}},
  {A101} ({2013}).
  
\item {{Bergin}, E.}, {{Calvet}, N.},
  {{D'Alessio}, P.} \& {{Herczeg}, G.~J.}
\newblock {{The Effects of UV Continuum and Ly{$\alpha$}
  Radiation on the Chemical Equilibrium of T Tauri Disks}}.
\newblock \emph{{Astrophys. J.L}} \textbf{{591}},
  {L159--L162} ({2003}).
  
\item {{Graninger}, D.~M.}, {{Herbst}, E.},
  {{{\"O}berg}, K.~I.} \& {{Vasyunin}, A.~I.}
\newblock {{The HNC/HCN Ratio in Star-forming Regions}}.
\newblock \emph{{Astrophys. J.}} \textbf{{787}},
  {74} ({2014}).
\end{enumerate}

\newpage
\section*{Extended Data}

\renewcommand{\figurename}{Extended Data Figure}
\setcounter{figure}{0}   
\renewcommand{\tablename}{Extended Data Figure}
\setcounter{table}{0}   

\begin{table}[htb]
\caption{\sf Physical model for the disk of MWC~480 \label{tab:model}}
\begin{tabular}{l c}
\hline
Parameters & Values \\
\multicolumn{2}{c}{Stellar properties} \\
\hline
Estimated distance: d (pc) & 140 \\
Stellar mass: M$_*$ (M$_\odot$)& 1.8 \\
\hline
\multicolumn{2}{c}{Disk structure properties} \\
\hline
Disk mass: M$_d$(M$_\odot$) & 0.18  \\
Characteristic radius: R$_c$(AU) & 81 \\
Outer cut-off radius (AU) & 100\\
Scale height: $H_{\rm 100AU}$ (AU)&16\\
Flaring index: $h$ &1.25\\
Density power-law index: $\gamma$&0.75\\
Midplane temperature: T$_{\rm 100 AU}$ (K) &23\\
Temperature power-law index: $q$ &0.5\\
Vertical temperature gradient index: $\beta$ &1.5\\
\hline
\multicolumn{2}{c}{Disk geometric and kinematic properties} \\
\hline
Inclination: $i$ (deg) & 37 \\
Systemic velocity: V$_{LSR}$(km s$^{-1}$) & 5.0 \\
Turbulent line width: v$_{turb}$(km s$^{-1}$) &0.05 \\
Position angle: P.A.(deg) & 58 \\
\hline
\end{tabular}
\end{table}

\begin{figure}
\includegraphics[width=0.6\textwidth]{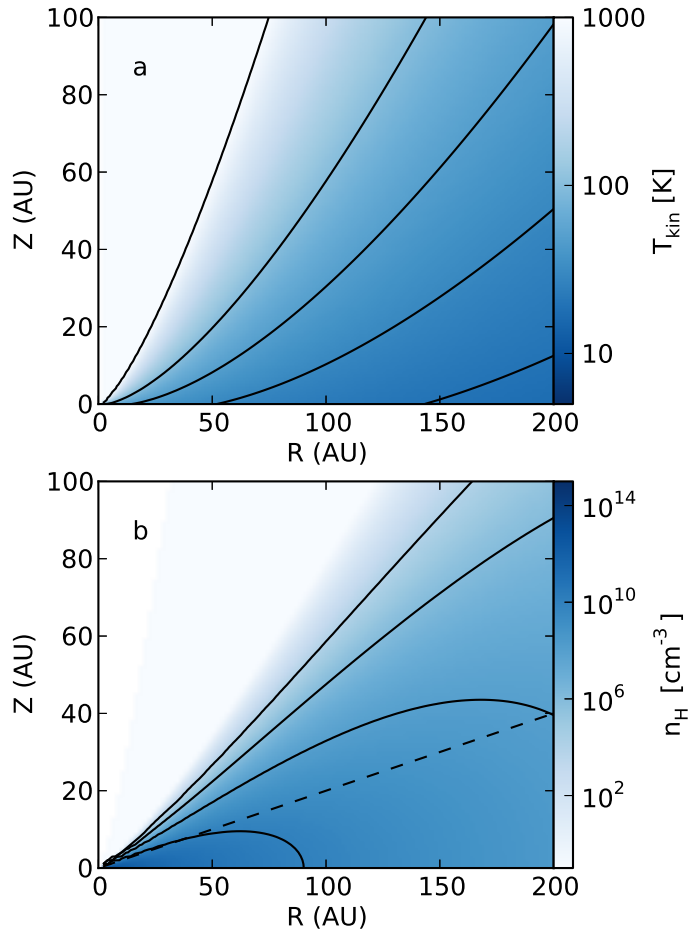}
\caption{\textbf{Model of the physical structure of the MWC 480 protoplanetary disk.} a, radial and vertical disk temperature profile (color: see color scale on the right, contours: 20,30,50,100, and 1000~K) with important temperatures marked with contours. b, radial and vertical density profile (color: see color scale on the right, contours: 10$^{10}$, 10$^8$, 10$^6$  and 10$^4$ cm$^{-3}$). The disk height over radius of 0.2 is marked as well. \label{fig_s1}}
\end{figure}

\begin{figure}
\includegraphics[width=1.0\textwidth]{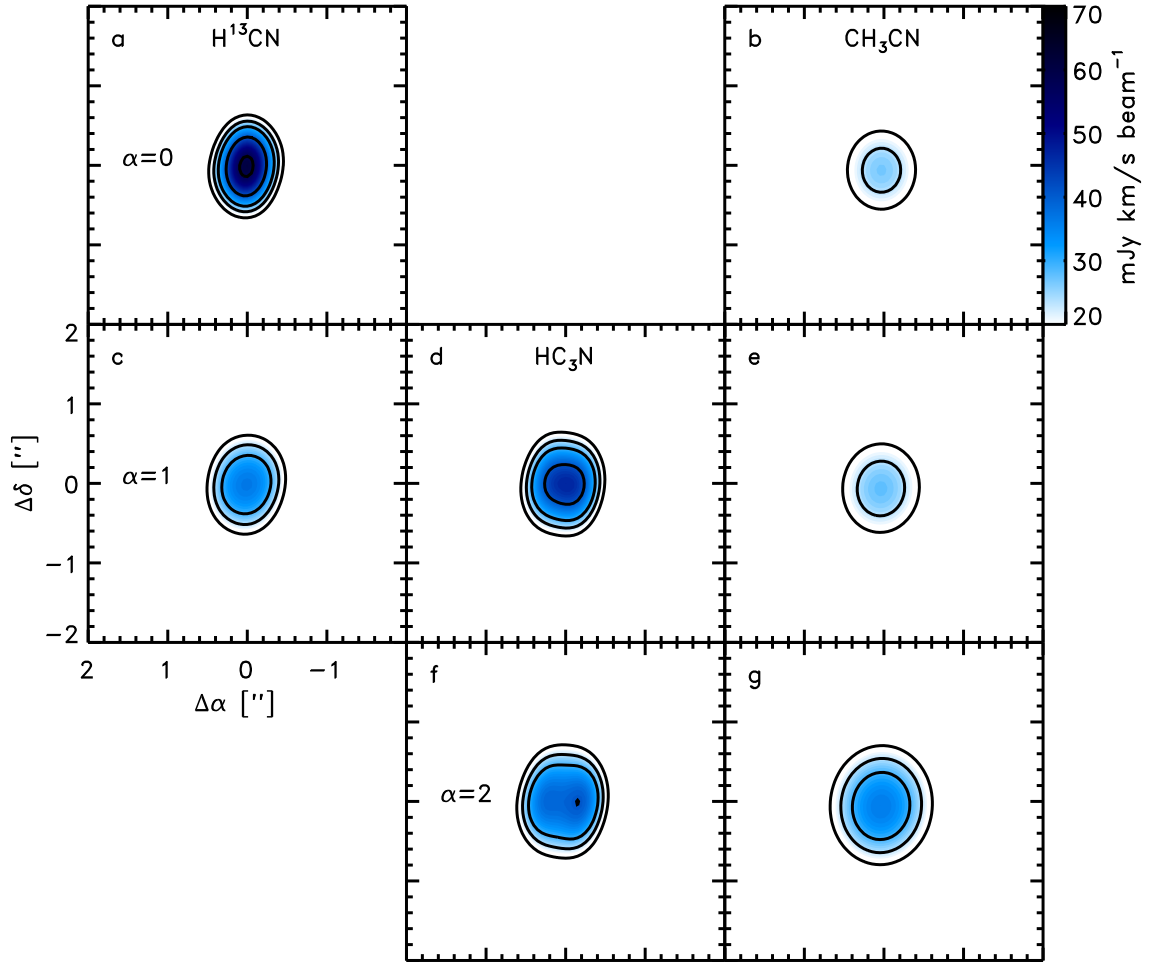}
\caption{\textbf{Synthetic observations of H$^{13}$CN, HC$_3$N and CH$_3$CN.} The models are based on best-fit to data for different choices of $\alpha$. a--g, integrated emission maps (colour: see colour scale on the right). Black contours are the observed [3,4,5,7,10]$\sigma$ in Fig. 1. The synthesized beam is shown in the bottom left corner of each panel. Note the change in emission profile between $\alpha=1$ and 2 for HC$_3$N.\label{fig_s2}}
\end{figure}

\begin{figure}
\includegraphics{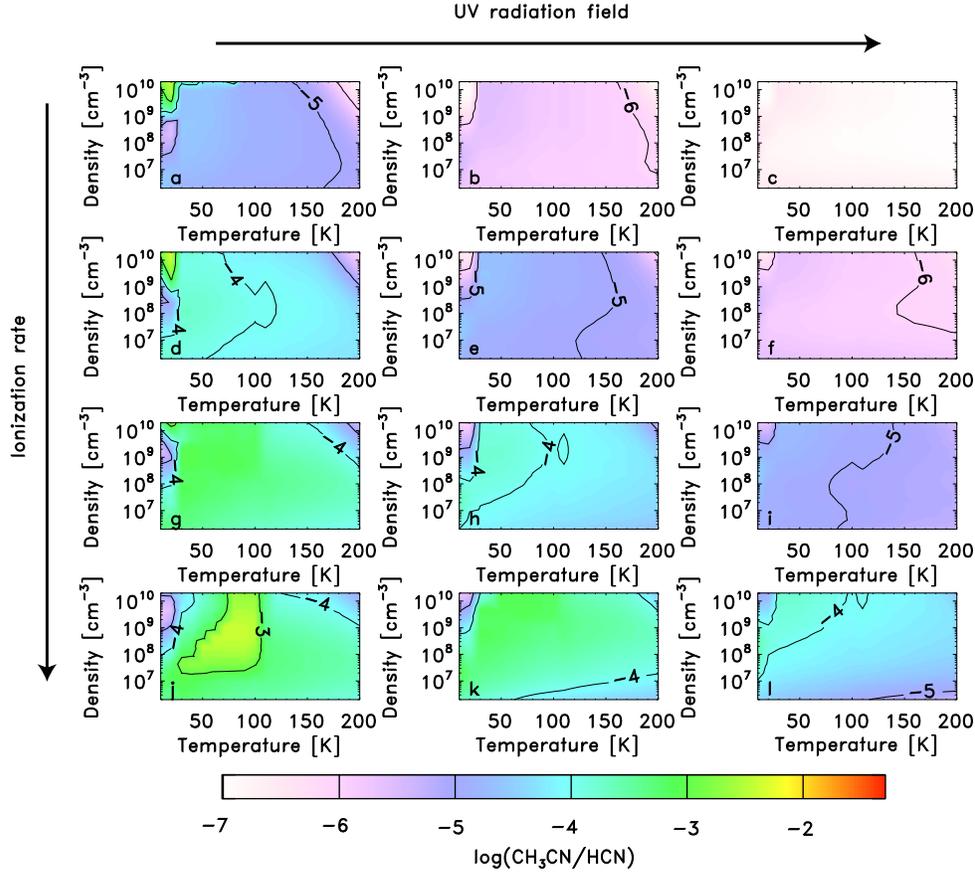}
\caption{\textbf{Models of gaseous CH$_3$CN/HCN abundance ratios under different physical conditions.} a--l, the CH$_3$CN/HCN abundance ratio on a logarithmic scale (colour: see colour scale on the bottom). UV radiation flux increases from left to right from $G_0 = 1$ (a,d,g,j) to $G_0 = 10$ (b,e,h,k)  to $G_0 = 100$ (c,f,i,l), where $G_0$ is scaling factor in multiples of the local interstellar radiation field. The ionization rate of H$_2$ increases from top to bottom from 10$^{-17}$ s$^{-1}$ (a--c) to 10$^{-16}$ s$^{-1}$ (d--f) to 10$^{-15}$ s$^{-1}$ (g--i) to 10$^{-14}$ s$^{-1}$ (j--l).\label{fig_s3}}
\end{figure}

\begin{figure}
\includegraphics[width=1.0\textwidth]{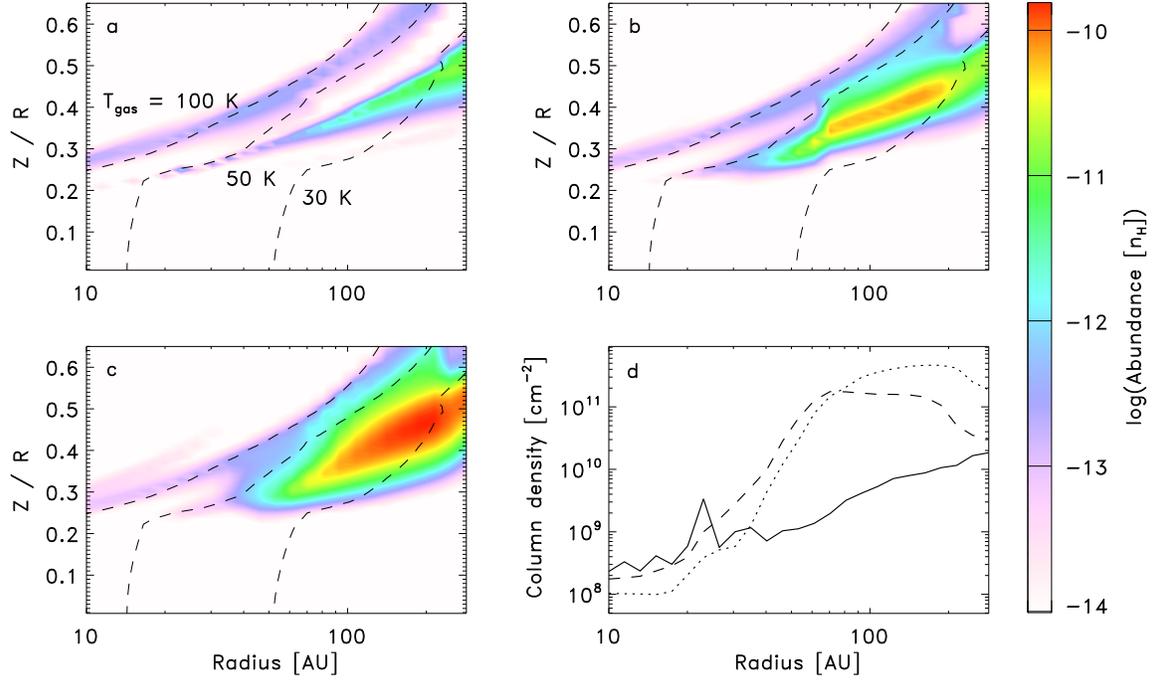}
\caption{\textbf{Models of gaseous CH$_3$CN in disks with and without turbulent diffusion.} a,  the abundance of CH$_3$CN (colour: see colour scale on the right) as function of disk radius ($R$) and height scaled by the radius ($Z/R$) in a model without turbulence. The dashed lines indicate gas temperatures of [30,50,100] K. b--c as a, but in disk models that include turbulence parameterized by $\alpha_z = 10^{-3}$ (b) and $\alpha_z = 10^{-2}$ (c). d, the vertically integrated column density of CH$_3$CN from a--c (solid line: $\alpha_z = 0$, dashed line: $\alpha_z = 10^{-3}$, dotted line: $\alpha_z = 10^{-2}$).}
\end{figure}

\begin{figure}
\includegraphics[width=1.0\textwidth]{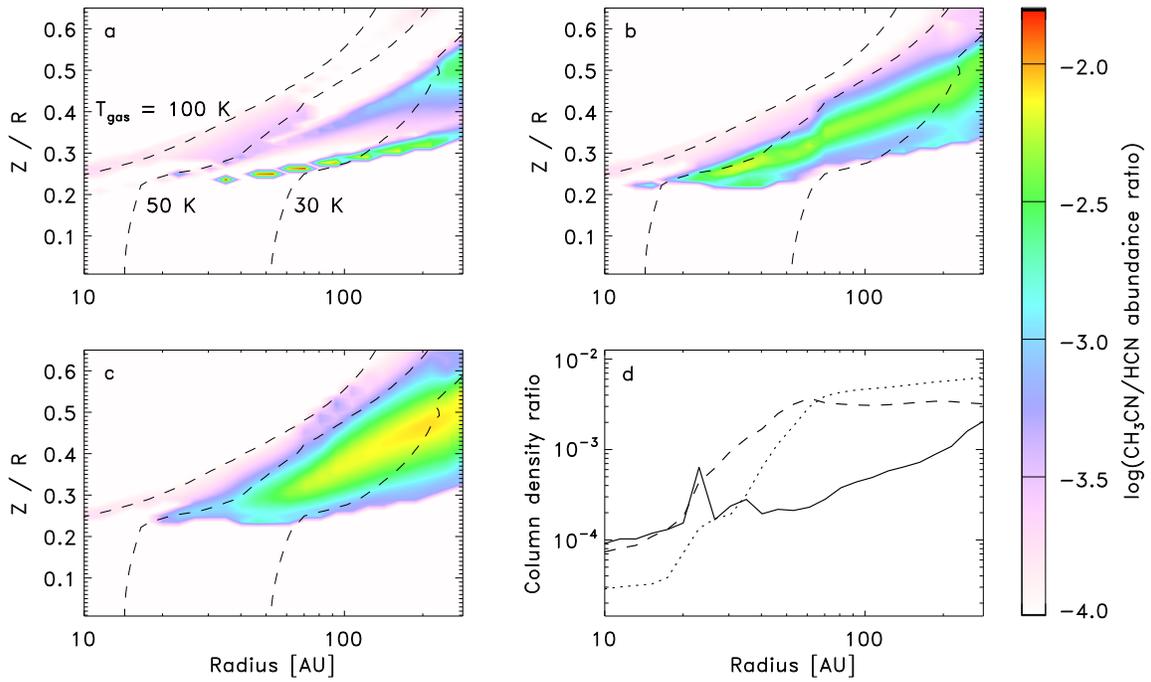}
\caption{\textbf{Models of gaseous CH$_3$CN/HCN ratios in disks with and without turbulent diffusion.}  a--d, as in Extended Data Fig. 4.}
\end{figure}

\begin{figure}
\includegraphics[width=1.0\textwidth]{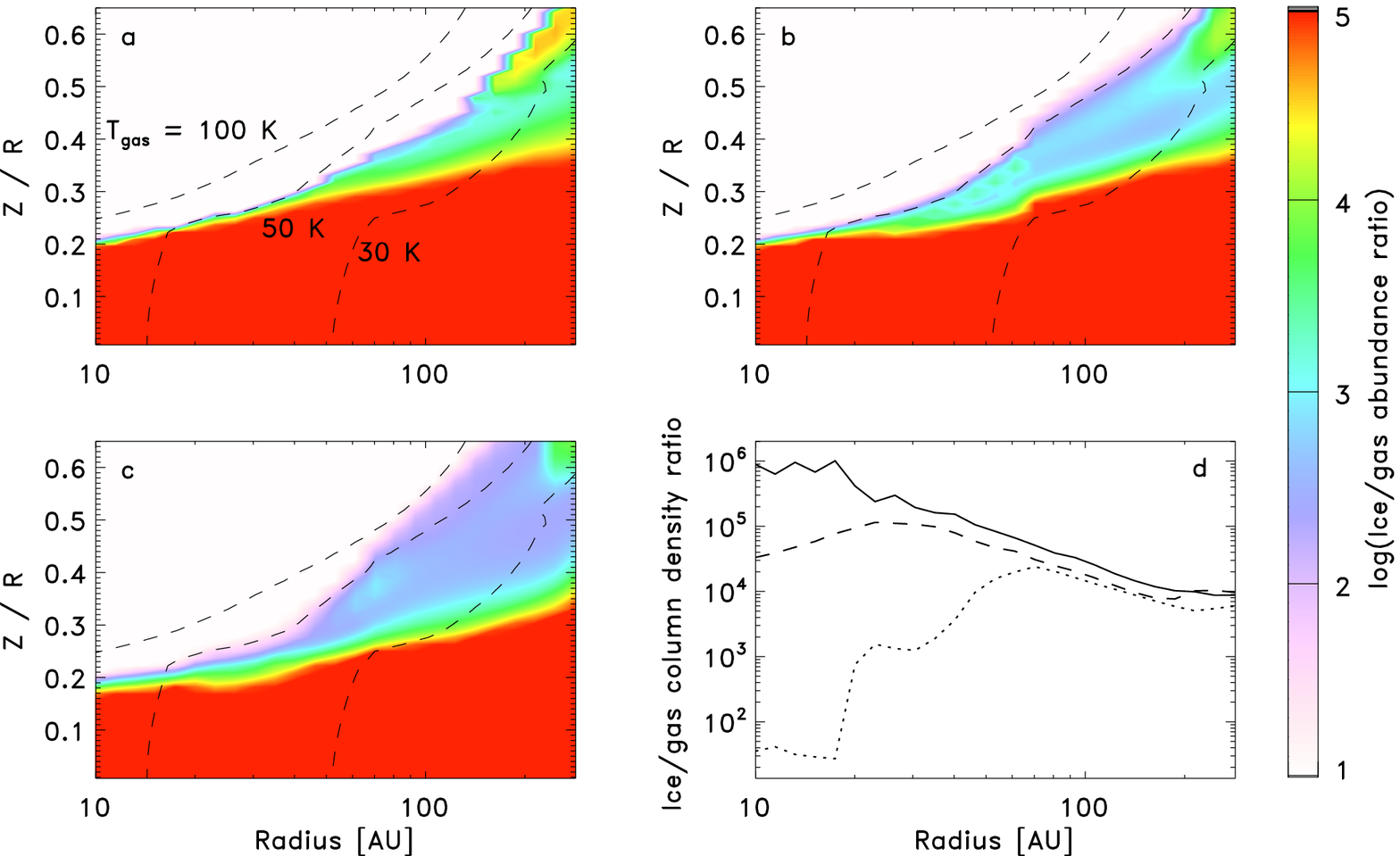}
\caption{\textbf{Models of gas-to-ice ratios of HCN in disks with and without turbulent diffusion.}  a--d, as in Extended Data Fig. 4.}
\end{figure}

\begin{figure}
\includegraphics[width=1.0\textwidth]{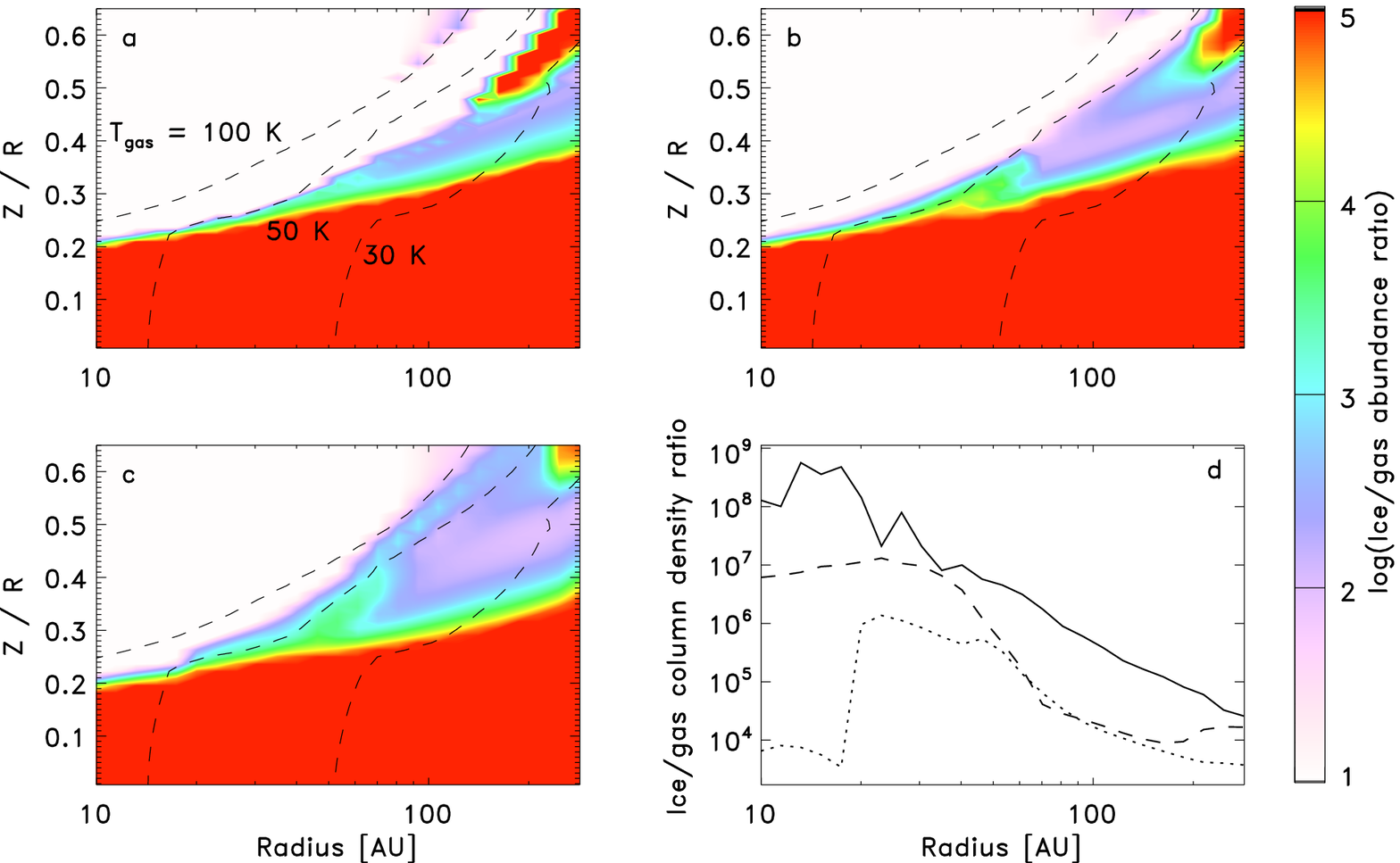}
\caption{\textbf{Models of gas-to-ice ratios of CH$_3$CN in disks with and without turbulent diffusion.}  a--d, as in Extended Data Fig. 4.}
\end{figure}

\end{document}